\documentclass[12pt]{article}

\usepackage{amsmath,amssymb,amsthm}
\usepackage{mathrsfs}
\usepackage{physics}

\usepackage{graphicx}
\usepackage{float}
\usepackage[font=footnotesize,labelfont=bf]{caption}

\usepackage{booktabs}
\usepackage{multirow}

\usepackage{algorithm}
\usepackage{algpseudocode}

\usepackage{enumitem}
\usepackage{xcolor}
\usepackage{textcomp}
\usepackage{fancyhdr}

\usepackage[title]{appendix}

\theoremstyle{plain}

\theoremstyle{definition}

\theoremstyle{remark}

\title{Stochastic Analysis of Fifth-Order KdV Soliton in Damping Regime and Reduction to Painlev\'e Second Equation}

\author{
Irfan Mahmood$^{1}$,
Adeena Iqbal$^{1}$,
Sohail Mumtaz$^{2}$\\[1ex]
$^{1}$Center for High Energy Physics, University of the Punjab,\\
Quaid-e-Azam Campus, Lahore 54590, Pakistan\\
$^{2}$Department of Physics, Seoul University, South Korea
}

\date{}

\begin{document}
\maketitle

% =============================
% Abstract
% =============================
\begin{abstract}
This work presents a stochastic analysis of fifth-order KdV soliton momentum distribution in a damping regime.
An explicit representation of the soliton momentum associated with amplitude variation is derived in terms of
a random time function in the presence of dissipation.
Statistical interpretations of soliton propagation modes, amplitude fluctuations, and amplification are analyzed
within a $\delta$-correlated Gaussian random framework.
Graphical results obtained using Python illustrate the physical insight into amplitude fluctuation and energy flow.
Finally, under a dominant approximation, the nonlinear momentum evolution equation is shown to reduce to the
Painlev\'e second equation, a well-known integrable model appearing in diverse physical systems.
\end{abstract}

\section*{Introduction}
The statistical analysis of integrable systems with their soliton solutions has received substantial attention from mathematical and physical perspectives because of their wide applications. The emergence of soliton equations in applied sciences has opened new research avenues to explore statistical aspects of their multi-soliton solutions as the system of $n$-particles with random variable functions. In this context few systems have been studied in \cite{21}-\cite{FK} and results are found very intersting from applications point of views. The soliton like solutions of various integrable systems have been studied in different frameworks \cite{1}-\cite{7} with the generalizations of their $N$ solutions in classical and non-commutative settings. The statistical interpretations as stochastical analysis, the random matrix  interpretation  of soliton energy and amplitude distribution associated with  random variables emerged  as quite interesting research areas from the implementations point of views in nonlinear material engineering. In this work, we discuss some statistical dynamics of the momentum dependent amplitude of  soliton  of fifth-order KdV-type Kaup–Kupershmidt first (KK-I) equation which is also connected to Kaup–Kupershmidt second (KK-II) equation through scale transformations \cite{1}.
These equations play a role of  integrable models in nonlinear control theory, fluid flow probes in viscous elastic medium, and nuclear dynamics in background of interactions \cite{8, 9}. Additionally, these equations \cite{10, 11} have been applied in study of fractional integrals and nonlocal symmetric examples  which combine the mathematical and physical perspectives of complex problems. The two version of  Kaup–Kupershmidt \cite{12, 13} have been successfully
used to describe the propagation of magneto-ion acoustic waves in plasma, the study of ocean waves, and in non-linear optics \cite{14, 15}.
Here the standard KK-I equation 
\begin{equation}
u_t + 4u^2 u_x - \frac{75}{2} u_x u_{xx} - 15u u_{xxx} + u_{xxxxx} = 0,
\label{eq:KK1}
\end{equation}
is modified to describe the propagation of soliton with amplitude variations in damping regime with random variable leading term in following 
\begin{equation}
u_t + 4u^2 u_x - \frac{75}{2} u_x u_{xx} - 15u u_{xxx} + u_{xxxxx} =\alpha(t)u+\beta \frac{\partial^2{u}} {\partial x^2},
\label{2}
\end{equation}
where $\alpha(t)$ is the time dependent leading coefficient and $\beta$ is a damping parameter.
Here we consider the KK-I equation due to its implementation as a simpler model for investigating the integrable features of higher-order KdV system and analysis on it may cover cover a large variety of higher order equations in the KdV hierarchy. It is also linked to other higher order equations for example, it is connected to  the KK-II equation via scaling transformations.  Initially, Kaup \cite{16} presented its solitary wave solution and  later on, two and three-soliton solutions were found by using the Hirota method. Further  Wronskian generalizations of its  higher-order solitons are recently  constructed  in \cite{2} and also admitted as completely integrable system \cite{12, 13} as  possesses  canonical form,  Lax representation  and Hamiltonian structure. 
The work presented in \cite{17} regarding the  excitations of oscillatory phenomenon  with parameter fluctuations has attracted attention in soliton theory to investigate the soliton amplitude response with variation of random function. In this context, the solitons associated with  nonlinear Schrodinger (NLS) equation  and the KdV equation the stochastic parametric resonances \cite{18} have been explored for time dependent amplitudes  under the various condition of dissipation. \\  Further to strengthen the statistical analysis of solitons to treat as system of particles, the $N$-soliton  dynamics are  deformed as system of $N$ particles with  random fields \cite{19, 20}. In one of these works, it has been shown that parametric action excites the internal degrees of freedom of solitons \cite{20} which enriches this research avenue from the physical point of view. In addition to these, for  well known NLS and KdV solitons, the various  amplification existence of their amplitudes in damping framework with  stationary behavior has been studied by \cite{21}. Here we investigate the soliton deriving force in terms of time rate of change of momentum which pushes the soliton in presence of dissipation. Further, we explicitly calculate  soliton momentum  in terms of a random variable function of time and dissipation parameter. Finally, we analyze the statistical interpretations of various modes of the propagation of  soliton with momentum fluctuation ,   amplitude variations in framework of $\delta$-correlated random process for Gaussian random function. The results are also presented with graphical  interpretations using python.\\ 
Finally in section $4$, we establish a connection of momentum function, stationary energy flow profiles, with Painlev\'e second equation. This link is made possible  under a dominant approximation for nonlinear momentum evolution equation which reveals that momentum distribution functions are the solutions of one Painlev\'e second equation, more about the significance of this link has been discussed in section $4$.

\section*{The soliton driving force: Time rate of change of momentum of one-soliton:}
In \cite{2}, the basic one-soliton solution of Eq. (\ref{eq:KK1}) without deformations is calculated as below
\begin{equation}\label{3}
u = A(t) \sech^2 \left[ \chi(t) (x + Vt) \right],
\end{equation}
here 
$A(t) = -2k(t)^{3/2}$ is momentum dependent amplitude with  $\quad V = \frac{k^2}{4}$ and $\quad \chi(t) = k^{\frac{1}{3}}$.
\begin{figure}[h!]
	\centering
\includegraphics[width=0.30\linewidth, height=0.25\textheight]{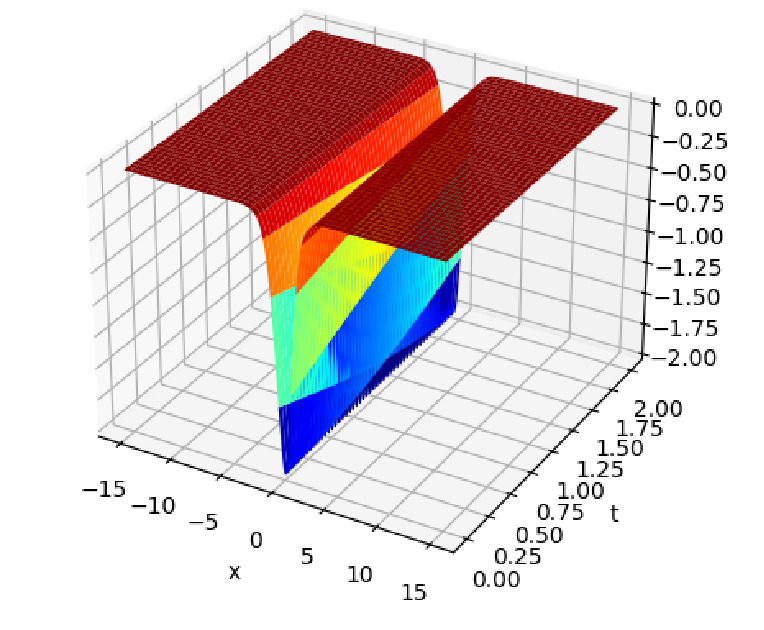}~~~~~~~~
\includegraphics[width=0.30\linewidth, height=0.25\textheight]{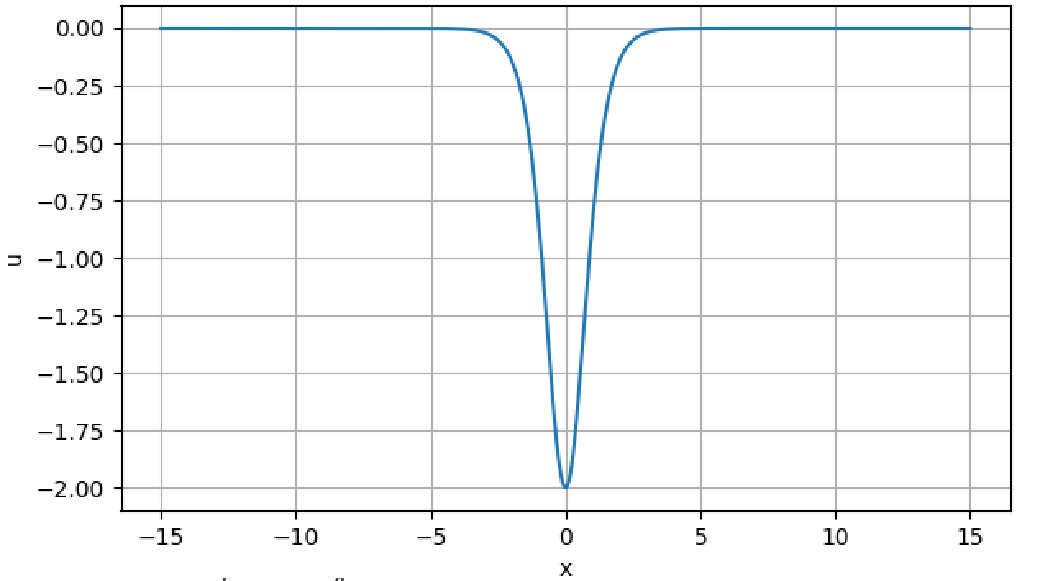}~~~~~~~~
\includegraphics[width=0.30\linewidth, height=0.25\textheight]{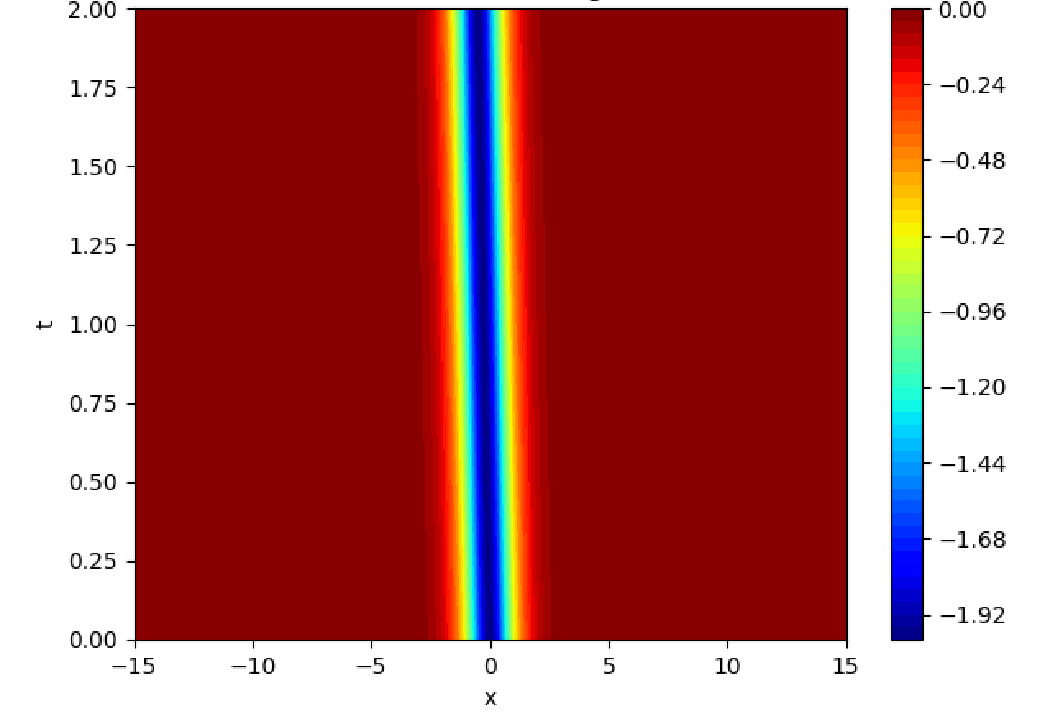}~~~~~~~~
    \\
(a) 3D \ \ \ \ \ \ \ \ \ \ \ \ \ \ \ \ \ \ \ \ \ \ \ \ \ \ \ \ \ \  \ \ \ \ \ (b) 2D  \ \ \ \ \ \ \ \ \ \ \ \ \ \ \ \ \ \ \ \ \ \ \ \ \ \ \ \ \ \  (c) contour
\\
\caption{(a) Three dimensional profile of single soliton. (b) Two dimensional perspective of single soliton solution. (c) contour representation of single soliton.}\label{fig1}
\end{figure}
\begin{figure}[h!]
	\centering
\includegraphics[width=0.30\linewidth, height=0.25\textheight]{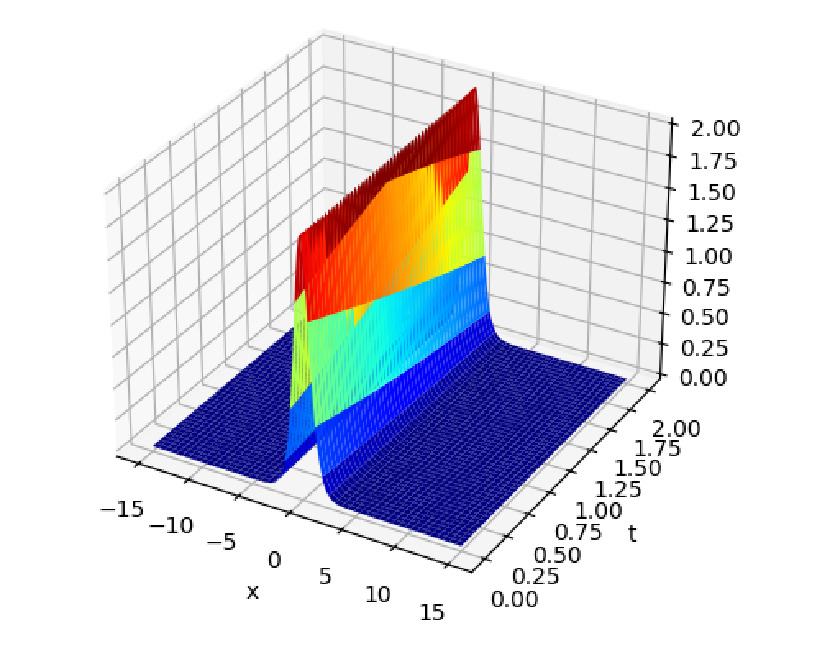}~~~~~~~~
\includegraphics[width=0.30\linewidth, height=0.25\textheight]{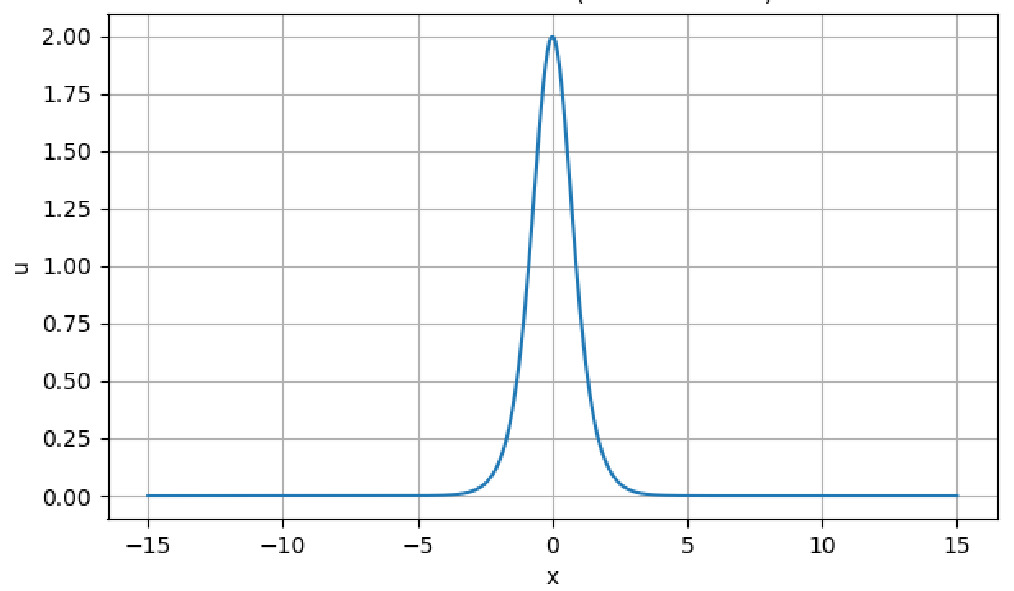}~~~~~~~~
\includegraphics[width=0.30\linewidth, height=0.25\textheight]{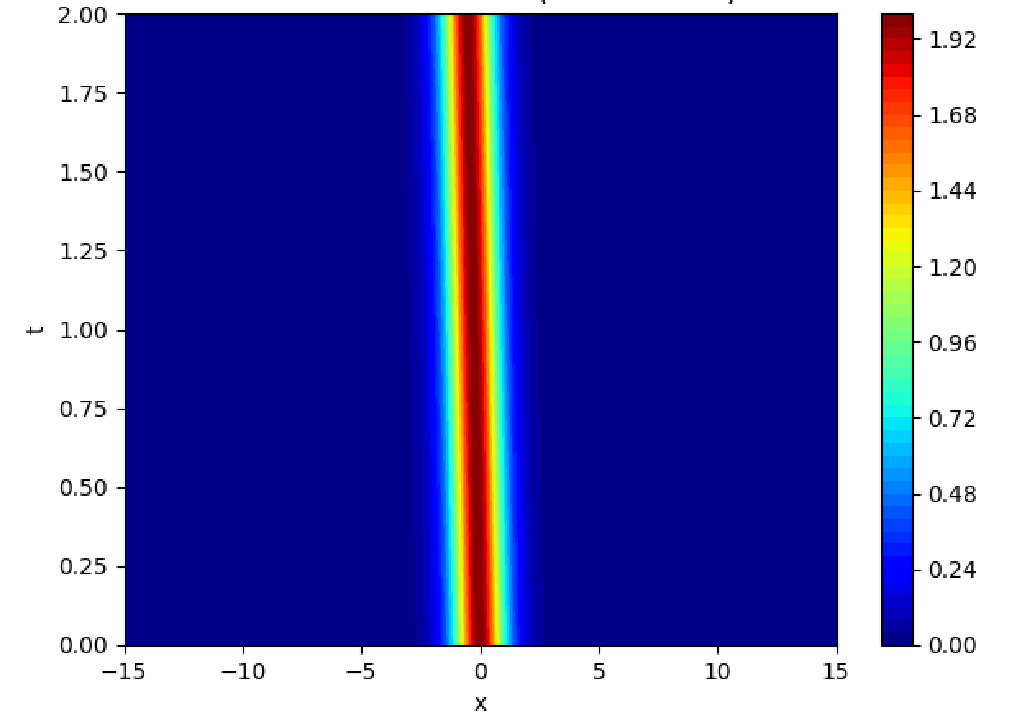}~~~~~~~~
    \\
(a) 3D \ \ \ \ \ \ \ \ \ \ \ \ \ \ \ \ \ \ \ \ \ \ \ \ \ \ \ \ \ \  \ \ \ \ \ (b) 2D  \ \ \ \ \ \ \ \ \ \ \ \ \ \ \ \ \ \ \ \ \ \ \ \ \ \ \ \ \ \ \ (c) contour
\\
\caption{(a) Three dimensional profile of single soliton. (b) Two dimensional perspective of single soliton solution.(c) contour representation of single soliton.}\label{fig2}
\end{figure}
Now  the momentum of one soliton (\ref{3}) under the boundary conditions as $x 	
\rightarrow ±\infty $, $u(x,t) \rightarrow 0$, to behave like a particle, over the whole space is given by $P = \int_{-\infty}^{+\infty} u^2(x,t) \, dx$. Now by defining  \(\eta = x(t) (x + \chi(t))\) then that  momentum can be calculated as 
\[
P = \frac{A_0^2}{\chi(t)} \int_{-\infty}^{+\infty} \sech^4 \eta \, d\eta = \frac{A_0^2}{\chi(t)} \int_{-\infty}^{+\infty} \sech^4 \eta \, d\eta = -\frac{8}{3} k, 
\]
and its  time derivation can be calculated as  
 \begin{equation}\label{Q1}
\frac{dP}{dt} = -\frac{8}{3} \frac{dk}{dt}.
\end{equation}
Now let calculate the $\frac{dP}{dt}$ through an alternative by using equation of motion 
 (\ref{2}) in presence dissipation and random time variable coefficient, as below  
\begin{equation}\label{a}
\frac{dP}{dt} = \frac{2}{\chi(t)} \int_{-\infty}^{+\infty} u u_t \, d\eta ,
\end{equation}
and after simplification, we get resulting expression with non-vanishing terms in following form 
\begin{equation}\label{P1}
\frac{dP}{dt} = - \frac{2 \alpha(t)}{\chi(t)} \int_{-\infty}^{+\infty} u^2 \, d\eta - \frac{2 \beta^2}{\chi(t)} \chi^2(t) \int_{-\infty}^{+\infty} u u_{\eta \eta} \,d\eta .
\end{equation}
The first integral in above expression  can easily be calculated as
\begin{equation}\label{b}
-\frac{2 \alpha(t)}{\chi(t)} \int_{-\infty}^{+\infty} u^2 \, d\eta = - \frac{8}{3} \alpha \frac{A^2}{\chi(t)} = - \frac{8}{3} \alpha(t) k ,
\end{equation}
and second integral can be solved by parts 
\begin{equation} \label{c}
 \begin{aligned}
-\frac{2 \beta}{ \chi(t)}  \chi^2(t) \int_{-\infty}^{+\infty} u_\eta u_{\eta\eta} \, d\eta &= -2 \beta \chi(t) A^2 \left[\int_{-\infty}^{+\infty}\sech^6\eta\, d\eta - \int_{-\infty}^{+\infty} \sech^4\eta\, d\eta \right] \\
&= -2 \beta A^2 \chi^2(t) \left[ \frac{16}{15} - \frac{4}{3} \right] = \frac{32}{15} \beta k^2. 
\end{aligned}
\end{equation}
Now from (\ref{P1}),  we get
\begin{equation}\label{Q2}
\frac{dP}{dt} = -\frac{8}{3} \alpha(t) k + \frac{32}{15} \beta k^2     
\end{equation}
After comparing (\ref{Q1}) and (\ref{Q2}) we obtain
\begin{equation}
\frac{dk}{dt} = \alpha(t) k - \frac{4}{5} \beta k^2.
\end{equation}\label{T3}
The above differential equation is nonlinear riccati equation, which  can be solved by integrating factor method for $k(t)$, time dependent momentum, for soliton amplitude in following form
\begin{equation}
k(t) = k_0 e^{\int \alpha(t) dt} \left[ 1 + \frac{4}{5}\beta F(t) \right]^{-1}, 
\end{equation}\label{T4}
where $F(t) = \int e^{\int \alpha(t) dt} dt$, now by using binomial expansion  for $\beta<<1$, we get
\begin{equation} \label{T5}
k(t)=k_{0}e^{\int \alpha(t)dt}[1-\frac{4}{5}\beta k_{0}\int e^{\int \alpha(t)dt}dt].
\end{equation}
In following section, we we show that above equation (\ref{T5}) will play a crucial role to investigate  stochastical analysis of soliton amplitude  variations under various approximations as  its  amplification in damping regimes and  some resonances are observed for different values of random Gaussian function on time scale.   
\section*{ Stochastical  analysis with random time function parameter}
Let the parameter of the random time function \(\alpha(t)\) be defined as  $\alpha(t) = \alpha_0 + \overline{\alpha}(t)$ where
\begin{itemize}
\item \(\alpha_0 = \langle \alpha \rangle\): is the  mean value
\item \(\overline{\alpha}(t)\): is the parametric random Gaussian function
\end{itemize}
Now in the framework of $\delta$-correlated random process, we define
\[ 
\langle \overline{\alpha(t)} \rangle = 0, \quad \langle \overline{\alpha}(t_1) \overline{\alpha}(t_2) \rangle = 2\alpha^2 \delta(t_1 - t_2), 
\]	
taking $\alpha_{0}=0$, then from  expression (\ref{T5}) we may construct  
\begin{equation}
<k^2>=k_{0}^2 e^{2 \sigma^2 t}[1-\frac{4}{5}\beta k_{0}\int e^{2 \sigma^2 t}dt], 
\end{equation}
or
\begin{equation}
<k^2>=k_{0}^2 e^{2 \sigma^2 t}[1-\frac{4}{5}\frac{\beta k_0}{\sigma^2}(e^{2 \sigma^2 t}-1)],
\end{equation}
the time flow of $<k^2>$ at different values of $\sigma^2$ has been show below.

\begin{figure}[h!]
	\centering
\includegraphics[width=0.45\linewidth, height=0.3\textheight]{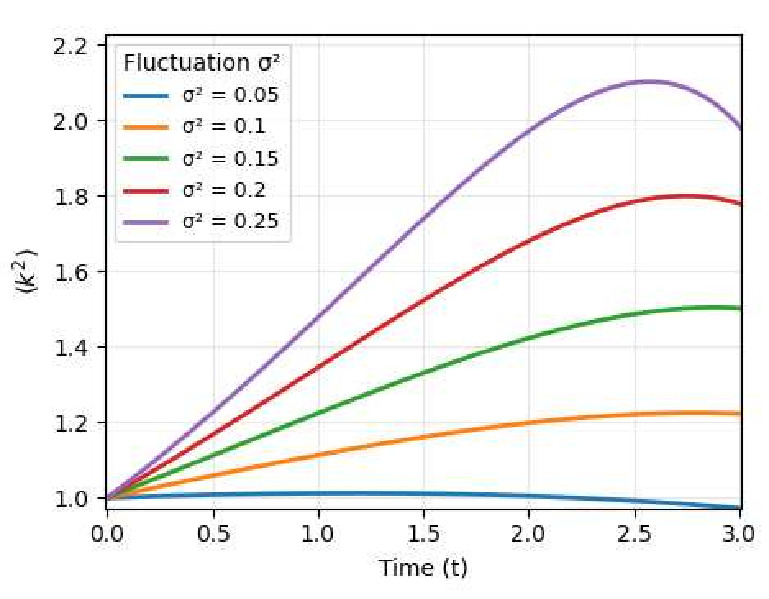}~~~~~~~~
    
\caption{ Here graph shows the variations in amplitudes on time scale for different values of $\sigma^2$ while pulse propagating through damping medium.}\label{fig3}
\end{figure}
\newpage
If $\beta=0$, without damping, we have  
\begin{equation}
<k^2>=k_0^2 e^{2\sigma^2 t}.
\end{equation}
\begin{figure}[h!]
	\centering
\includegraphics[width=0.4\linewidth, height=0.3\textheight]{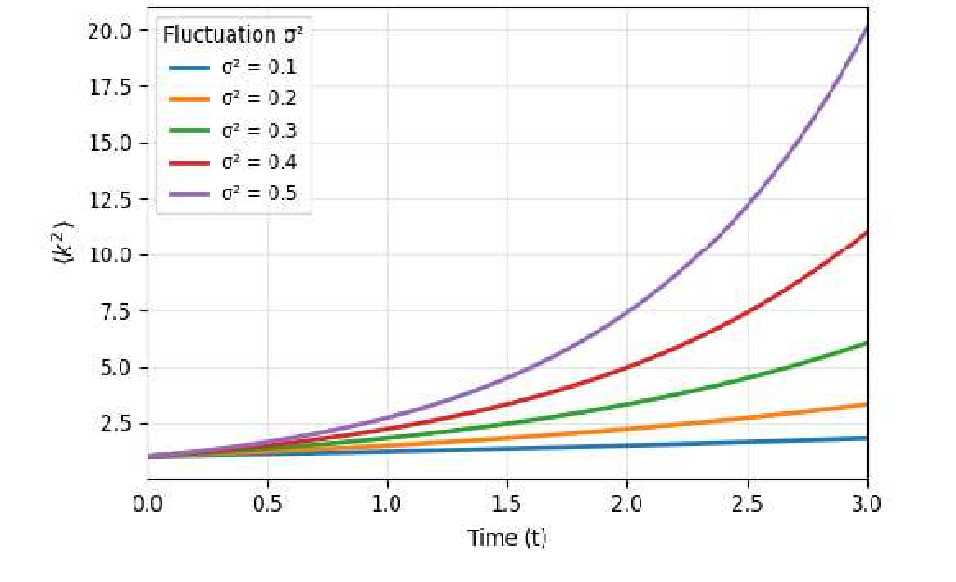}~~~~~~~~

\caption{Here graph shows the variations in amplitudes on time scale for different values of $\sigma^2$ while pulse propagating with zero dissipation.}\label{fig4}
\end{figure}
Here have we considered the time evolution of amplitudes without damping effects, the amplification of amplitude at larger values of  $\sigma$ is entirely exponential but at low values grows very slow. The growth of  amplitude in  time for larger values $\sigma$  sharply increases and  becomes uncontrolled.  
\newpage
If $\sigma^2 t<1$, we obtain family of straight lines originating from single point of intercept on  vertical axis $<k^2>$ as below
\begin{equation}
<k^2>=k_{0}^2 [1+2(\sigma^2-\frac{4}{5}\beta k_0)t]
\end{equation}
\begin{figure}[h!]
	\centering
\includegraphics[width=0.45\linewidth, height=0.3\textheight]{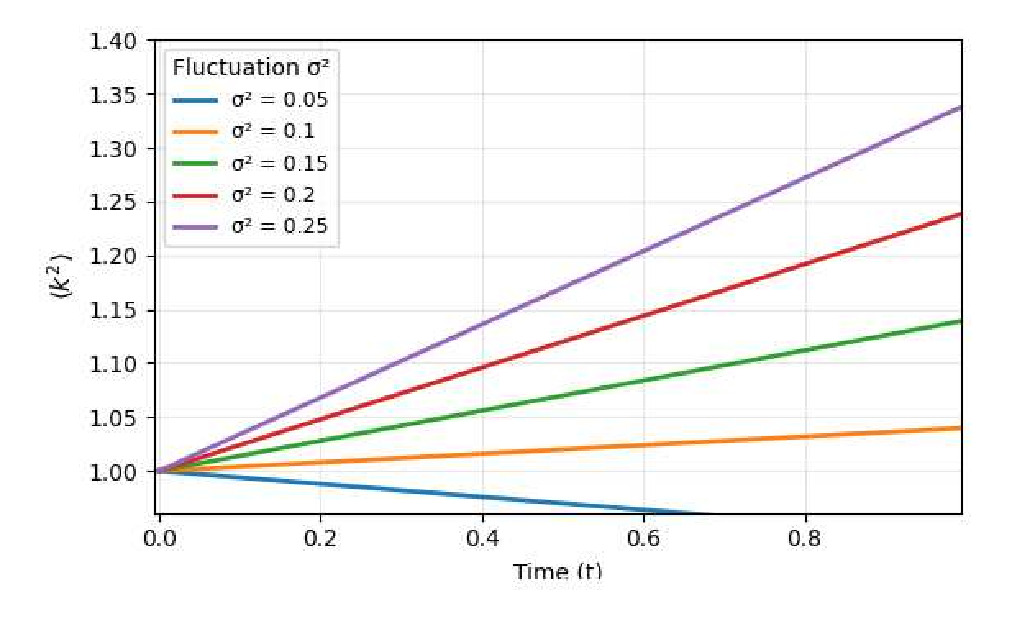}~~~~~~~~
    
\caption{Here, the graph shows the variations in amplitudes for $\sigma^2 t<1$ along the straight lines in the damping regime on the time scale. }\label{fig5}
\end{figure}
\section*{Momentum distribution as Painlev\'e second solutions }
For very small damping \( \beta <<< 1 \) we have  
\[ 
k(t) = k_0 e^{\int \alpha (t) d(t)} , 
\]  
and we get  
\begin{equation} \label{r1}
\frac{g(t)}{k(t)} = \alpha (t),
\end{equation} 
let here \( g(t) = \dot k(t)= \frac{dk}{dt} \) grows very slow near $t_0$ , then from Taylor series  
$g(t) = g(t_0) + g'(t_0) t + ... Negl.$
  
then from ( \ref{r1})  
\[
\alpha (t) = \frac{g(t_0) + \dot g(t_0)t}{k(t)},
\]  
and now  
\[
\frac{dk}{dt} = g(t_0) + \dot g(t_0)t + \lambda k^2,
\]  
by taking \( n = g(t_0),\; m = \dot g(t_0) \)  
\[
\frac{dk}{dt} = n + mt + \lambda k^2.
\]  
Now once time derivation of above equation yields ,  
\[
\frac{d^2k}{dt^2} = m + 2\lambda k \frac{
\dot dk}{dt} = m + 2\lambda k\left(n + mt + \lambda k^2\right),
\]
or
\[
\frac{d^2k}{dt^2} = m + 2\lambda k (n + mt) + 2\lambda^2 k^3, 
\]

Let Substitute  $n + mt = \tau$, we get
  
\[
m^2 \frac{d^2k}{d\tau^2} = m + 2\lambda k \tau^2 + 2\lambda^2 k^3,
\]  
or  
\[
\frac{d^2k}{d\tau ^2} = \frac{1}{m} + \frac{2\lambda}{m^2} k \tau + \frac{2\lambda^2}{m^2} k^3, 
\]

 Now introducing  scaling $  k \rightarrow \left( \frac{2m}{\lambda^2} \right)^{1/3}q, \quad 
\gamma \rightarrow \left( \frac{m^2}{2\tau} \right)^{1/3} z$ in above expression, we obtain the Painlev\'e second equation as below
  \begin{equation}\label{r2}
\frac{d^2k}{dz^2} = qz + 2q^3 + \delta ,  
\end{equation}
here  the parameter $\delta $ and above Painlev\'e second equation \cite{PP}  is a well know integrable model appears in various domains of physics and mathematics. That equation possesses the rational solutions, the Yablonskii-Vorob'ev polynomial, for integral values \cite{AL, AP}and admits the Airy's type solutions \cite{HAR}for odd half values of parameter.  The Painlev\'e second equation \ref{r2}  has been acknowledged as completely  integrable model and implemented to explain the stationary electric field fluctuations \cite{NK} in nonlinear semiconductors and its advanced versions \cite{6, 7, MUH}as quantum-noncommuative have also been studied with its detreminantal  solutions.  
\section*{Graphical Interpretation}
Figures $1, 2$ manifest soliton profiles associated with  higher order KdV type Kaup–Kupershmidt equation in  two and three dimensional with their contour graphs. In figure $3$ it has been shown that how the amplitude is behaving for different values of $\sigma^2 $ on time scale for various cases. At smaller value of $\sigma^2 =0.05$ amplitude gains energy for very short time and the declines immediately. But for the large of $\sigma^2 =0.25$ amplitude rises rapidly, pulse pushed by force with positive time rate of change in its momentum, then reaches the maximum value resonance occurs and then intimidate damping becomes dominant.  The green curve at $\sigma^2 =0.15$ includes very interspersing behavior from application point of view, this shows we can attenuate the amplitude, propagating pulse, at maximum energy value for comparatively a long time in damping regime.   Figure $4$, shows the explanation increase in amplitudes, uncontrolled response in the absence of dissipation and the presences of guiding term with time random variable. Figure $5$ represent fluctuation of amplitudes is purely linear under the  approximate $\sigma^2 t<1$. From the graphical analysis the results presented in figure $3$ are very substantial from applications point of views. 
\section*{Conclusion}
Here we have presented  stochastic analysis of one-soliton momentum distribution in damping regime. We have explored  the dynamics of soliton with the interpretations of its amplitude variations, amplifications, resonances and sustaining of its profile in different damping regimes $\delta$-correlated random process. As we have considered higher order , the fifth order KdV like equation, these computations can be extended to higher order members in KdV hierarchy. The main motive is to extend this analysis for the semi-discrete and fully discrete  analog  of the Kaup–Kupershmidt equation. This finding on discrete analog will be towards more generalization of  stochastic analysis of multi-solitons on lattice and reducible to continuum region for multi-dimensional cases.  Our results , specially presented in figure $3$  are very interesting from the application point fo view to attenuate  the energy pulse for the desirable maximum value for a long time in form of  soliton in damping regime. This may be applicable in technology of nonlinear optical fibre and energy propagation  in magneto-hydrodynamics background of damping effects with parametric attenuation at desire levels.

\end{document}